\documentstyle[prc,preprint,tighten,aps]{revtex}

\title{\bf Neutron charge radius and the Dirac equation}

\author{M.\ Bawin\thanks{Electronic-mail address: michel.bawin@ulg.ac.be}}
\address{Universit\'{e} de Li\`{e}ge, Institut de Physique B5, Sart
Tilman, 4000 Li\`{e}ge 1, Belgium  }

\author{S.\ A.\ Coon\thanks{Electronic-mail address: coon@nmsu.edu}}
\address{Physics Department, New Mexico State University,
         Las Cruces, NM  88003, USA}

\begin{document}

\maketitle

\begin{abstract} We consider the Dirac equation for a finite-size
neutron in an external electric field.  We explicitly incorporate
Dirac-Pauli form factors into the Dirac equation.  After a
non-relativistic reduction, the Darwin-Foldy term is cancelled by a
contribution from the Dirac form factor, so that the only coefficient
of the external field charge density is $\textstyle{\frac{e}{6}}
r^2_{En}$, {\em i. e.} the root mean square radius associated with the
electric Sachs form factor . Our result is similar to a recent result
of Isgur, and reconciles two apparently conflicting viewpoints about
the use of the Dirac equation for the description of nucleons.
\end{abstract}


\vskip20pt

The description of a spin half particle ( e.g., neutron, proton) which has
internal structure by means of the Dirac equation remains controversial.
The seemingly innocuous assumption  that the neutron
behaves like a Dirac particle has
led to quite a controversy over the meaning of the charge radius of the
neutron.   This quantity is measured in low energy neutron-atom
scattering and thus it is linked to the electrical polarizability;
sometimes both are extracted from the same experiment~\cite{Koester}.
The disparate values~\cite{Leeb,Riehs} in the literature are cited as, on
the one hand, consistent with chiral bag models~\cite{Byrne} and, on
the other hand, quite inconsistent with these same models of the
substructure of the neutron~\cite{Alex}.  This dispute is based on the
denial or acceptance of the Foldy term~\cite{Foldy} which arises in the use
of  a Dirac-Pauli equation for the interaction of the neutron
 with an external electromagnetic  field~\cite{Riska}. The issue was
recently revived by Isgur in the context of the valence quark
model~\cite{Isgur}.  For the valence quark model, Isgur concludes that
the Foldy term  is cancelled by a contribution from the
Dirac form factor.  We will discuss here the form this cancellation
takes  when the Dirac equation is used in conjunction with
electromagnetic form factors of the Dirac-Pauli form. That is, a
Foldy-Wouthuysen type analysis of the Dirac equation yields a result
similar to that of  Isgur, but it makes
 no explicit reference to the nature of the internal structure of the
neutron.  The result follows only from the fact that the neutron is not
a point particle  and the assumption that the neutron is described by
the  Dirac equation with a form factor description of its non-zero
size.

	Consider the Dirac Hamiltonian for a neutron of mass m in an external
electric  field ${\bf E}$.  We include  form factors to
describe the composite nature of the neutron.  The natural form factors
of the Dirac equation are $F_1(t)$ and $F_2(t)$, defined by the momentum
space charged-matter four-current  density $j^+_\kappa(x)$ in the presence
of an electromagnetic field and for positive energy scattered particles:
\begin{equation}
  e\langle {\bf p}'|j^+_\mu(0)|{\bf p}\rangle = e\bar{u}({\bf p}')
      [F_1(t) \gamma_\mu + F_2(t)i \sigma_{\mu\nu} q^\nu
      /2m ] u({\bf p}) \;,
\end{equation}
where $t = q^2 =(p - p')^2$ and $\bar{u}({\bf p}')$ and $u({\bf p})$ are free particle Dirac
spinors. We employ the contemporary
 definitions of {\em dimensionless} form factors (see, for example,
 Ref.\cite{Scadronbook}), so that $F_1(0) = 1$ for a charged and $F_1(0)
 = 0$ for a neutral particle. Then $F_2(0) \equiv \kappa
 = \textstyle{\frac{1}{2}}(g-2)$ is the dimensionless, anomalous magnetic
 moment of a  spin-$\textstyle{\frac{1}{2}}$ particle~\cite{units}.
 The four momentum transferred from the photon to the spin
$\textstyle{\frac{1}{2}}$ particle is $q = p - p'$.    In the Breit frame,
necessary for assigning a coordinate space meaning to the nucleon
electromagnetic form factors,  $ {\bf p} + {\bf p}' = 0$ so that $q_0 =
0$, and    $t = (p - p')^2 = -{\bf q}\,^2$.  The  electric and magnetic
Sachs form factors are defined by
\begin{eqnarray}
	G_E(t) &=& F_1(t) + \frac{t}{4m^2} F_2(t) \nonumber \\
	G_M(t) &=& F_1(t) +F_2(t)  \label{eq:Sachs}
\end{eqnarray}
In the Breit frame, the  Sachs form factors have  simple
interpretations as the spatial Fourier transforms of the nucleon's
charge and magnetization distributions~\cite{Sachs}.  We need only
\begin{equation}
	\rho_N({\bf r}) = \left (\frac{1}{2\pi}\right)^3
	\int d^3 {\bf q}\, e^{+i {\bf q}\cdot { \bf r}}
		G_E(-{\bf q}^2)\;,  \label{eq:rhoGE}
\end{equation}		
such that the normalization $\int  d^3 {\bf r}\,\rho_N({\bf r})$ is one for the
proton and zero for the neutron.

 We will concentrate on the small momentum transfer aspects of this
 problem, so write the usual expansions
 of $F_1$ and $F_2$:
\begin{equation}
F_i(t)=F_i(0)+tF_i'(0)+ \cdots \;\;\;, i=1,2\;\;,  \label{eq:Fexpand}
\end{equation}
so that
\begin{equation}
G'_E(t) = F'_1(t) + \frac{F_2(t)}{4m^2} + \frac{t}{4m^2} F'_2(t)\;,
   \label{eq:Gprime}
\end{equation}
and
\begin{equation}
G'_E(0) = F'_1(0) + \frac{F_2(0)}{4m^2} = F'_1(0) + \frac{\kappa}{4m^2}\;,
	\label{eq:Gprime0}
\end{equation}
where $\kappa$ {\em is} the (dimensionless) magnetic moment  of
the neutron. However,  one defines the root mean square radius of the
charge distribution of the neutron by means of the inverse Fourier
transform
\begin{equation}
G_E(-{\bf q}^2) = \int d^3 r e^{-i {\bf q}\cdot {\bf r}} \rho({\bf r})
  = -\frac{1}{6} r^2_{En}{\bf q}^2 + \cdots\;\;,
\end{equation}
with the corresponding
definition $ G_E(-{\bf q}^2) = 1 - \frac{1}{6} r^2_{Ep}{\bf q}^2  + \cdots $
 for the proton.  Since
 $G'_E(t=0) = -\textstyle{\frac{\partial{G_E({\bf q}^2=0)}}{\partial{{\bf
 q}^2}}}$, equation (\ref{eq:Gprime0}) becomes
\begin{equation}
 F'_1(0) \equiv \frac{\partial{F_1({\bf q}^2=0)}}{\partial{{\bf q}^2}}
 = -\frac{1}{6} r^2_{En} + \frac{\kappa}{4m^2}\;.  \label{eq:F'}
\end{equation}

After this discussion of the electromagnetic form factors
 we are ready to write down the Dirac Hamiltonian operator as:
\begin{equation}
	H = \mbox{\boldmath $\alpha$} \cdot {\bf p}
	+ \beta m + i
	\frac{eF_2}{2m}\beta\mbox{\boldmath $\alpha$}
	\cdot{\bf E} +
		eF_1V\;,   \label{eq:Ham}
\end{equation}
where $F_1$ and $F_2$ are, respectively, the neutron Dirac and Pauli
 form factors, and  $V$ is the  potential associated with
 ${\bf E}$.
   Note that the use of Breit variables then leads in a natural way to
using the Coulomb gauge for $V({\bf q})$.
>From equations (\ref{eq:Gprime}) and (\ref{eq:Gprime0}) it should
 be apparent that only $F_2(0) = \kappa$ plays a role
in the charge radius of the neutron, so we rewrite (\ref{eq:Ham}) as

\begin{equation}
	H = \mbox{\boldmath $\alpha$} \cdot {\bf p} +\beta m + i
	\frac{e  \kappa}{2m}\beta\mbox{\boldmath
	$\alpha$}\cdot{\bf E} +
		eF_1V\;   \label{eq:Ham2}
\end{equation}
The latter (\ref{eq:Ham2}) is then a simple extension of the model used
in our discussion of the neutron polarizability~\cite{BCpol,musign}.
Replacement of the function $F_2$ by its limit does not lead to a loss
of generality but does
simplify the non-relativistic reduction of the Dirac equation
$H\psi = E\psi$, to which we now turn.  We work in coordinate space
where
\begin{equation}
	\tilde{V}({\bf r}) \equiv eF_1V = e \left (\frac{1}{2\pi}\right)^3
	\int  d^3 {\bf q}\, e^{i {\bf q}\cdot
	{\bf r}} F_1
	(-{\bf q}\,^2)V({\bf q})\,,    \label{eq:efv}
\end{equation}
and we retain the symbol ${\bf p}$ for the momentum operator $-i
\mbox{\boldmath $\nabla$}$ for clarity.  We begin the non-relativistic limit by
squaring $H$ of (\ref{eq:Ham2}) to get~\cite{BS}
\begin{equation}
   \{ (E-\tilde{V})^2 - {\bf p}^2 -m^2 -\frac{e^2\kappa^2 {\bf E}^2}{4m^2}
   + i \mbox{\boldmath $\alpha$} \cdot \mbox{\boldmath $\nabla$}\tilde{V}
   + \frac{e\kappa }{2m}\beta (\mbox{\boldmath $\nabla$}\cdot {\bf E})
   + 2\frac{e}{2m}\mbox{\boldmath $\kappa$}\cdot ({\bf E} \times {\bf p})
       \} \psi = 0\;.  \label{eq:exact}
\end{equation}
This equation is exact.  Now we set $E \sim m$, $\beta\sim 1$,  $\tilde{V}
\ll m$, and employ the approximation $\mbox{\boldmath $\alpha$} \sim
\textstyle{\frac{ \bf p}{2m}}$ discussed in equation (12.7) of Bethe and
Salpeter's treatise~\cite{BS}.  We then get
\begin{eqnarray}
	\{ \frac{{\bf p}^2}{2m} - \frac{{e\bf p}}{2m^2}\cdot
({\bf E} \times \mbox{\boldmath $\kappa$} ) +
\frac{e^2 \kappa^2 {\bf E}^2}{8m^3} -
\frac{e\kappa }{4m^2} (\mbox{\boldmath $\nabla$}\cdot {\bf E}) & & \nonumber \\
+ \tilde{V} + \frac{\mbox{\boldmath $\nabla$}^2}{8m^2} \tilde{V}
 \} \psi &
= E\psi\;.   \label{eq:nonrel}
\end{eqnarray}
Here we have separated the nonrelativistic terms   new to this
Hamiltonian ( {\em i.e.}, those with
$\tilde{V}$ which vanish for a point Dirac particle) from those already displayed in equation (4) of
Ref.\cite{BCpol}  obtained by a Foldy-Wouthuysen
transformation.  We have checked that the systematic Foldy-Wouthuysen
coordinate space procedure, which is more tedious and requires, in our
case, switching back and forth between the coordinate and momentum
representations of the Dirac equation, yields the same result
(\ref{eq:nonrel}). After completion of this work, we learned of the
Hamiltonian obtained by McVoy and van Hove~\cite{MvH} using the Foldy-Wouthuysen
procedure and checked that it is equivalent to equation
(\ref{eq:nonrel}).

In order to exhibit explicitly the $r^2_{En}$ in the non-relativistic
equation (\ref{eq:nonrel}) we
must return to momentum space using the  definition
$\tilde{V}({\bf q}) =  e  F_1(-{\bf q}\,^2)V({\bf q})$
implicit in (\ref{eq:efv}).
Now we use the low ${\bf q}^2$ expansion of $F_1$ (see
(\ref{eq:Fexpand}))  and write

\begin{eqnarray}
	\lefteqn{\tilde{V}({\bf q}) = e\int d^3{\bf q}\,e^{i {\bf q}\cdot
	{\bf r}} F_1
	(-{\bf q}\,^2)V({\bf q})
	= e\int d^3{\bf q}\,e^{i {\bf q}\cdot{\bf r}}
	  [0 + {\bf q}^2 F'_1(0) + \cdots] V({\bf q}) } \nonumber \\
  & &	\approx - e F'_1(0) \mbox{\boldmath $\nabla$}^2 V({\bf r})
	= eF'_1(0)\mbox{\boldmath $\nabla$}\cdot {\bf E}
	=-e \left ( \frac{1}{6} r^2_{En} -
	\frac{\kappa}{4m^2}\right ) \mbox{\boldmath $\nabla$}\cdot {\bf
	E}\;,
		\label{eq:vtilde}
\end{eqnarray}
 where we have used (\ref{eq:F'}) and remind the reader that
 $ {\bf E}({\bf r}) =
   - \mbox{\boldmath $\nabla$}V({\bf r})$.

By substituting (\ref{eq:vtilde}) into (\ref{eq:nonrel}), we observe that
the Foldy term (the term  $\frac{e\kappa}{4m^2}$) in (\ref{eq:nonrel}) is
now cancelled by a contribution from the {\em Dirac} form factor,
leaving $\textstyle{\frac{e}{6}} r^2_{En}$ as the only coefficient of
the external field charge density.  This
is  similar to Isgur's result~\cite{Isgur}.   Our result, however, is
independent of any definite  quark substructure of the neutron. The
neutron just has to have a form factor.  Furthermore, this analysis is
in keeping with the philosophy that the basic  equations for the
neutron should be expressed in terms of a Dirac  Hamiltonian, while the
physical picture emerges from a nonrelativistic  reduction which only
contains $r^2_{En}$. In our
analysis, we express the basic interaction in terms of a Dirac  form factor
$F_i$, but acknowledge that the physics lies in Sachs form  factors
$G_E$ and $G_M$.

In order to avoid a misunderstanding of our result, we note that it is
the $\kappa$ term buried in the Sach's form factor $G_E$
(\ref{eq:Sachs})
which contributes the most to the neutron charge radius, even after the
 cancellation of the Foldy term.  The dominance of the $\kappa$ term
in the neutron charge radius has been noted before by
Friar~\cite{Friar}.

It is of some interest to note that the left hand side of
(\ref{eq:nonrel}) also contains a
term $\textstyle{\frac{\mbox{\boldmath $\nabla$}^2}{8m^2}} \tilde{V}$,
which has the form of a Darwin term, but this time associated with the
{\em neutron} charge distribution rather than that of the external
field. In particular, for $V = -e/r$ this term is simply
\begin{equation}
	\frac{\mbox{\boldmath $\nabla$}^2}{8m^2} \tilde{V}({\bf r}) =
	\frac{+e}{8m^2} \rho^1_{En}({\bf r}) \;,
\end{equation}
where
\begin{equation}
\rho^1_{En}({\bf r}) =\left (\frac{1}{2\pi}\right)^3
	\int  d^3 {\bf q}\, e^{i {\bf q}\cdot
	{\bf r}} F_1
	(-{\bf q}\,^2)\;.
\end{equation}

Finally, let us note that our analysis and conclusion are in complete
agreement with the analysis of low-energy Compton scattering from the
nucleon by L'vov~\cite{Lvov}.  In that work Dirac form factors appear
in an effective Lagrangian for the $\gamma n$ interaction, but the
relation (\ref{eq:F'}) is (in effect)  used in order to exhibit the low energy
Compton amplitude expressed in terms of physical observables, as
required by low energy theorems~\cite{Shekhter,Autrans}. Our result
also reconciles the two apparently  conflicting viewpoints
\cite{Byrne,Alex} about the use of the Dirac equation for the
description of nucleons.

\section*{Acknowledgements}
The work of M.B. was supported by the National Fund for Scientific
Research, Belgium and that of S.A.C. by NSF grant PHY-9722122.  We are
very grateful to Dan-Olof Riska for informing us of Ref. \cite{MvH}.

\end{document}